\RequirePackage{etoolbox}
\csdef{input@path}{{style/}{graphics/}}
\documentclass[aoas,MSNbibl,nameyear,dvips]{arximspdf}
\makeatletter
   \@ifpackageloaded{graphicx}{}{\usepackage{graphicx}}
\makeatother
\usepackage{url,breakurl}

\doi{10.1214/14-AOAS803}
\volume{9}
\issue{1}
\pubyear{2015}
\firstpage{145}
\lastpage{165}
\docsubty{FLA}
\setattribute{copyright}{owner}{In the Public Domain}

\begin{document}
\begin{frontmatter}

\title{Continuous-time discrete-space models for animal~movement}
\runtitle{Discrete-space movement models}

\begin{aug}
\author[A]{\fnms{Ephraim M.}~\snm{Hanks}\corref{}\thanksref{M1}\ead[label=e1]{hanks@psu.edu}},
\author[B]{\fnms{Mevin B.}~\snm{Hooten}\thanksref{M2,M5}}\\
\and
\author[C]{\fnms{Mat W.}~\snm{Alldredge}\thanksref{M3}}
\runauthor{E. M. Hanks, M. B. Hooten and M. W. Alldredge}
\affiliation{Pennsylvania State University\thanksmark{M1},
U. S. Geological Survey, Colorado Cooperative Fish and Wildlife
Research Unit\thanksmark{M2},
Colorado State University\thanksmark{M5} and
Colorado Parks and Wildlife\thanksmark{M3}}
\address[A]{E. M. Hanks\\
Department of Statistics\\
Pennsylvania State University\\
325 Thomas Building\\
University Park, Pennsylvania 16802\\
USA\\
\printead{e1}}
\address[B]{M. B. Hooten\\
U.~S. Geological Survey\\
Colorado Cooperative Fish\\
\quad and Wildlife Research Unit\\
Department of Fish, Wildlife\\
\quad and Conservation Biology\\
Colorado State University\\
201 JVK Wagar Bldg.\\
Fort Collins, Colorado 80523\\
USA}
\address[C]{M. W. Alldredge\\
Colorado Parks and Wildlife\\
317 W. Prospect Rd.\\
Fort Collins, Colorado 80526\\
USA}
\end{aug}

%
\received{\smonth{12} \syear{2013}}
%
\revised{\smonth{11} \syear{2014}}

%
\begin{abstract}
The processes influencing animal movement and resource selection are
complex and varied. Past efforts to model behavioral changes over time
used Bayesian statistical models with variable parameter space, such
as reversible-jump Markov chain Monte Carlo approaches, which are
computationally \mbox{demanding} and inaccessible to many practitioners. We
present a continuous-time discrete-space (CTDS) model of animal
movement that
can be fit \mbox{using} standard generalized linear modeling (GLM) methods.
This CTDS approach allows for the joint modeling of location-based
as well as directional drivers of movement.
Changing behavior over time is modeled using a varying-coefficient
framework which maintains the computational simplicity of a GLM
approach, and variable selection is accomplished using a group lasso
penalty. We apply our approach to a study of two mountain lions
(\emph{Puma concolor}) in Colorado, USA.
\end{abstract}

%
\end{frontmatter}

\section{Introduction}\label{sec1}
Telemetry data have been used extensively in recent years to
study animal movement, space use and resource selection
[e.g., \citet{Johnson2011,Hanks2011PLoS,Fieberg2010}].
The simplest form of telemetry data consist of a time series of
remotely obtained spatial
locations of an animal. Typically, an animal or group of animals are
captured and fit with a tracking device (e.g., a collar with a GPS)
which records the animal's location at specified intervals.
The ease with which telemetry data are being collected is increasing,
leading to vast improvements in the number of animals being monitored, as
well as the temporal resolution at which telemetry locations are
obtained [\citet{Cagnacci2010}]. This combination can result in huge
amounts of telemetry
data on a single animal population under study.
Additionally, the processes driving animal movement are complex,
varied and changing over time. For example, animal behavior could be
driven by the local environment [e.g., \citet{Hooten2010b}], by
conspecifics or predator/prey
interactions [e.g., \citet{Merrill2010,Potts2014}], by internal states and
needs [e.g., \citet{Nathan2008}], or by memory [e.g., \citet
{VanMoorter2009}]. The
animal's response to each of
these drivers of movement is also likely to change over
time [e.g., Hanks et~al. (\citeauthor{Hanks2011PLoS}), \citet{McClintock2012}, \citet{Nathan2008}] as
animals respond to changing stimuli (e.g., dirunal cycles) or energy needs.


Examples of recent models for animal telemetry data include the
agent-based model of \citet{Hooten2010b}, the
velocity-based framework for modeling animal movement of
\citet{Hanks2011PLoS}, and the mechanistic approach of
\citet{McClintock2012}. These three approaches use
Markov chain Monte Carlo (MCMC) for inference, and both \citet
{Hanks2011PLoS} and
\citet{McClintock2012} allow for time-varying behavior by letting the
model parameter space vary, either through a reversible-jump Markov chain Monte Carlo approach [\citet{Green1995}] or the related
birth--death Markov chain Monte Carlo approach [\citet{Stephens2000}].
Such methods are computationally demanding and require the user to
tune the algorithm to ensure convergence. Our goal is to provide an
approach to modeling complex time-varying movement behavior that is both
scientifically useful and computationally tractable.

While telemetry data can be collected with relative ease at high
resolution, habitat covariates (i.e., landcover) are typically
available only in gridded form at a fixed resolution. Traditional
analyses that focus on modeling an animal's location often contain
redundant information because observations are close enough in time
that the
spatially available habitat data contains little information to model
the fine scale movement. Therefore, constructing an analysis with an
eye toward the habitat data scale holds promise for the future of
telemetry data.

In this manuscript, we present a continuous-time, discrete-space (CTDS)
model for animal movement which allows for
flexible modeling of an animal's response to drivers of movement in a
computationally efficient framework. We consider a Bayesian approach
to inference,
as well as a multiple-imputation approximation to the posterior distribution
of parameters in the movement model. Instead of a state-switching or
change-point model for
changing behavior over time, we adopt a time-varying coefficient
model.
We also allow for variable selection using a lasso penalty.
This CTDS approach is highly computationally efficient, requiring only
minutes or seconds to analyze movement paths that would
require hours using the approach of \citet{Hanks2011PLoS} or days using
the approach of \citet{Hooten2010b}, allowing the analysis of longer
movement paths and more complex behavior than has been previously possible.

In Section~\ref{sec2}, \emph{Continuous-time Markov chain
models for animal movement}, we describe the CTDS model
for animal movement and present a latent variable representation of
the model that allows for inference within a standard
generalized linear model (GLM) framework.
In Section~\ref{sec3}, \emph{Inference on CTDS model
parameters using telemetry data}, we present a
Bayesian approach for inference and describe the use of multiple imputation
[\citet{Rubin1987}] to approximate the posterior predictive distribution
of parameters in the CTDS model.
In Section~\ref{sec4},
\emph{Time-varying behavior and shrinkage estimation}, we use a
varying-coefficient approach to model changing behavior over
time, and use a lasso penalty for variable selection and
regularization. In Section~\ref{sec5},
\emph{Drivers of animal movement}, we discuss modeling potential
covariates in the CTDS framework. In Section~\ref{sec6}, \emph{Example}:
\emph{Mountain lions in Colorado}, we illustrate our approach through an
analysis of mountain lion (\emph{Puma concolor}) movement in Colorado,
USA. Finally, in Section~\ref{sec7}, \emph{Discussion}, we discuss possible
extensions to the CTDS approach.

\section{Continuous-time Markov chain models for animal movement}\label{sec2}

\begin{figure}[b]

\includegraphics{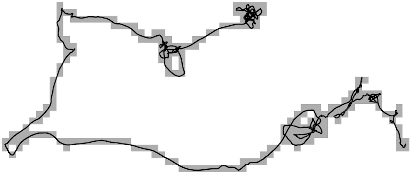}

\caption{Continuous-time continuous-space and continuous-time
discrete-space representations of an animal's movement path.}\label{fig1}
\end{figure}

Our goal is to specify a model of animal response to drivers of movement
that is flexible
and computationally efficient. We propose a continuous-time
Markov chain (CTMC) model for an animal's CTDS movement through a
discrete, gridded space (Figure~\ref{fig1}). We then present a latent variable
representation of a CTMC model that represents the CTMC as a
generalized linear model (GLM), allowing for inference in CTMCs in
general and CTDS movement models in particular to be
made using GLM theory and computation (e.g., iteratively reweighted
least squares optimization routines).


Let the study area be defined as a graph $(\mathbf{G},\mathbf{A})$
of $M$ spatial vertices $\mathbf{G}=(G_1,G_2,\ldots,G_M)$ connected by
``edges'' $\bolds\Lambda=\{\lambda_{ij}\dvtx i\sim j,i=1,\ldots,M\}$,
where $i\sim
j$ means that the nodes $G_i$ and $G_j$ are directly connected. For
example, in a gridded space each grid cell is a vertex (node) and the edges
connect each grid cell to its first-order neighbors (e.g., cells that
share an edge). In ecological studies, the spatial resolution of the grid
cells in $\mathbf{G}$ will often be determined by the resolution at which
environmental covariates that may drive animal movement and selection
are available. Discretizing an animal's path across the study area
amounts to studying movement at the spatial resolution of the
available landscape covariates.


An animal's continuous-time, discrete-space (CTDS) path $\tilde{\mathbf{S}}=(\mathbf{g},\bolds\tau)$
consists of a sequence of grid cells
$\mathbf{g}=(G_{i_1},G_{i_2},\ldots,G_{i_T})$ traversed by the
animal and the residence times
$\bolds\tau=(\tau_1,\tau_2,\ldots,\tau_T)$ in each grid cell.
The discrete-space representation $\tilde{\mathbf{S}}=(\mathbf
{g},\bolds\tau)$ of
the movement path allows us to use standard continuous-time Markov chain models to make inference about possible drivers of movement.

While we
will relax this assumption later to account for temporal
autocorrelation in movement behavior, we initially assume that the
$t$th observation $(G_{i_t},\tau_t)$ in the sequence is
independent of all other observations in the sequence. Under this
assumption, the likelihood of the sequence of transitions $\{(G_{i_t}
\rightarrow G_{i_{t+1}},\tau_t),t=1,2,\ldots,T\}$ is the product of
the likelihoods of each individual observation. We will focus on
modeling each transition $(G_{i_t} \rightarrow G_{i_{t+1}},\tau_t)$.

If an animal is in cell $G_{i_t}$ at time $t$, then define the rate of
transition from cell $G_{i_t}$ to a neighboring cell $G_{j_t}$ at time
$t$ as
%
\begin{equation}\label{equ1}
\lambda_{i_tj_t}(\bolds\beta)=\exp\bigl\{\mathbf{x}_{i_tj_t}'
\bolds\beta\bigr\},
\end{equation}
where $\mathbf{x}_{i_tj_t}$ is a vector containing covariates related
to drivers of movement specific to cells $G_{i_t}$ and $G_{j_t}$, and
$\bolds\beta$ is a vector of parameters that define how each of
the covariates in $\mathbf{x}_{i_tj_t}$ are correlated with animal
movement. The total transition rate $\lambda_{i_t}$ from cell
$G_{i_t}$ is the sum of the transition rates to all neighboring cells:
$\lambda_{i_t}(\bolds\beta)=\sum_{j_t \sim i_t}
\lambda_{i_tj_t}(\bolds\beta)$,
and the time $\tau_t$ that the animal resides in cell $G_{i_t}$ is
exponentially-distributed with rate parameter equal to the total
transition rate $\lambda_{i_t}(\bolds\beta)$:
%
\begin{equation}\label{equ2}
[\tau_t|\bolds\beta]=\lambda_{i_t}(\bolds\beta)
\exp \bigl\{-\tau_t \lambda_{i_t}(\bolds\beta) \bigr\}.
\end{equation}

When the animal transitions from cell $G_{i_t}$ to one of its neighbors,
the probability of transitioning to cell $G_{i_{t+1}}$, an event we denote
as $G_{i_t} \rightarrow G_{i_{t+1}}$, follows a multinomial
(categorical) distribution with
probability proportional to the transition rate $\lambda_{i_ti_{t+1}}$
to cell $G_{i_{t+1}}$:
%
\begin{equation}\label{equ3}
[G_{i_t} \rightarrow G_{i_{t+1}}|\bolds\beta]=
\frac{\lambda_{i_ti_{t+1}}(\bolds
\beta)}{\sum_{j_t\sim i_t} \lambda_{i_tj_t}(\bolds\beta)}=\frac
{\lambda_{i_ti_{t+1}}(\bolds\beta)}{\lambda_{i_t}(\bolds\beta)}.
\end{equation}
Under this formulation, the residence time and eventual
destination are independent events,
and the likelihood of the observation $(G_{i_t} \rightarrow
G_{i_{t+1}}, \tau_t)$
is the product of the likelihoods of its parts:
%
\begin{eqnarray}\label{equ4}
[G_{i_t} \rightarrow G_{i_{t+1}}, \tau_t|\bolds
\beta]&=&\frac{\lambda_{i_ti_{t+1}}(\bolds\beta
)}{\lambda_{i_t}(\bolds\beta)} \cdot\lambda_{i_t}(\bolds\beta) \exp \bigl
\{-\tau\lambda_{i_t}(\bolds\beta) \bigr\}
\nonumber\\[-8pt]\\[-8pt]\nonumber
&=&\lambda_{i_ti_{t+1}}(\bolds\beta)\exp \bigl\{-\tau_t
\lambda _{i_t}(\bolds\beta) \bigr\}.
\end{eqnarray}

\subsection{GLM representation of a continuous-time Markov chain}\label{sec2.1}
We now introduce a latent variable representation of the transition
process that is
equivalent to~(\ref{equ4}), but allows for inference within a GLM framework.
We note that this latent variable representation is applicable
to any continuous-time Markov chain model with transition rates
$\{\lambda_{i_tj_t}\}$ and provides a novel approach for inference to
this broad class of models. Representing a CTMC model as a GLM allows
us to analyze animal movement data using existing
computational methods for GLMs (i.e., estimation through
iteratively reweighted least squares). Computational efficiency is
important as our ability to collect long time series of
fine-resolution telemetry data increases.

For each $j_t$ such that $i_t\sim
j_t$, define $z_{i_tj_t}$ as
\[
z_{i_tj_t}=\cases{ 1, &\quad$G_{i_t} \rightarrow
G_{j_t}$,
\cr
0, &\quad o.w.}
\]
and let
%
\begin{equation}\label{equ5}
[z_{i_tj_t},\tau_t|\bolds\beta]\propto\lambda
_{i_tj_t}^{z_{i_tj_t}} \exp \bigl\{-\tau_t
\lambda_{i_tj_t}(\bolds\beta) \bigr\}.
\end{equation}
Then the product of $[z_{i_tj_t},\tau_t|\bolds\beta]$ over all
$j_t$ such
that $i_t\sim j_t$ is proportional to the likelihood (\ref{equ4}) of the
observed transition:
\begin{eqnarray*}
\prod_{j_t\dvtx i_t\sim j_t} [z_{i_tj_t},\tau_t|
\bolds\beta]&\propto&\prod_{j_t\dvtx i_t\sim
j_t}
\lambda_{i_tj_t}^{z_{i_tj_t}} \exp \bigl\{-\tau_t \lambda
_{i_tj_t}(\bolds\beta) \bigr\}
\\
&=&\lambda_{i_ti_{t+1}}(\bolds\beta)\exp \bigl\{-\tau_t
\lambda_{i_t}(\bolds\beta) \bigr\}\qquad\mbox{where }
G_{i_t} \rightarrow G_{i_{t+1}}
\\
&=& [G_{i_t} \rightarrow G_{i_{t+1}}, \tau_t|\bolds
\beta].
\end{eqnarray*}

The benefit of this latent variable representation is that the
likelihood of $z_{i_tj_t},\tau_t|\bolds\beta$ in (\ref{equ5}) is equivalent
to the
likelihood in a Poisson regression with the canonical
log link, where $z_{i_tj_t}$ are the observations and $\log(\tau_t)$ is
an offset or
exposure term. The likelihood of the entire continuous-time,
discrete-space path $\tilde{\mathbf{S}}=(\mathbf{g},\bolds\tau)$
can be written
as
%
\begin{equation}\label{equ6}
[\tilde{\mathbf{S}}|\bolds\beta] =[\mathbf{Z},\bolds\tau |\bolds
\beta] \propto\prod_{t=1}^T \prod
_{i_t \sim j_t} \bigl[\lambda_{i_t j_t}^{z_{i_tj_t}}(\bolds
\beta) \exp\bigl\{-\tau_t \lambda_{i_t j_t}(\bolds\beta)
\bigr\} \bigr],
\end{equation}
where $\mathbf{Z}=(\mathbf{z}_1,\ldots,\mathbf{z}_T)'$ is a vector
containing the latent variables
$\mathbf{z}_i=(z_{i_1},\break z_{i_2}, \ldots,z_{i_K})'$ for each grid cell in
the discrete-space path.

\section{Inference on CTDS model parameters using telemetry data}\label{sec3}
We have proposed a CTMC model for animal movement that relies on
a complete continuous-time discrete-space (CTDS) movement path
$\tilde{\mathbf{S}}=(\mathbf{g},\bolds\tau)$. In practice,
telemetry data are
collected at a discrete set of time points. Let
$\mathbf{S}=\{\mathbf{s}(t),t=t_0,t_1,\ldots,t_T\}$ be the observed sequence
of time-referenced telemetry locations for an animal. We propose a
two-step procedure for inference on $\bolds\beta$ in which we
first obtain a posterior predictive distribution
$[\tilde{\mathbf{S}}|\mathbf{S}]$ of the CTDS path
conditioned on the observed telemetry data $\mathbf{S}$.
In a Bayesian framework, we specify a
Gaussian prior on $\bolds\beta$ such that
%
\begin{equation}\label{equ7}
\bolds\beta\sim N(\mathbf{0},\bolds\Sigma_\beta)
\end{equation}
and then the posterior predictive distribution of $\bolds\beta$
conditioned only on the telemetry data $\mathbf{S}$ is given by
%
\begin{equation}\label{equ8}
[\bolds\beta|\mathbf{S}] =\int_{{\mathcal{S}}}[\bolds\beta |
\tilde{\mathbf{S}}] [\tilde{\mathbf{S}}|\mathbf{S}]\,d\tilde{\mathbf{S}}.
\end{equation}
\citet{Hooten2010b} and \citet{Hanks2011PLoS} use composition sampling
to obtain samples from a similar posterior predictive distribution
by sampling
iteratively from $[\tilde{\mathbf{S}}|\mathbf{S}]$ and
$[\bolds\beta|\tilde{\mathbf{S}}]$. In addition to this approach
(which we will call a fully Bayesian approach),
we also consider approximate posterior predictive inference on
$\bolds\beta$ using multiple imputation
[\citet{Rubin1987}]. 

\subsection{Multiple imputation}\label{sec3.1}

In the multiple imputation literature
[e.g.,\break \citeauthor{Rubin1987} (\citeyear{Rubin1987,Rubin1996})], $\tilde{\mathbf{S}}$ is treated
as missing data, and the posterior predictive path
distribution $[\tilde{\mathbf{S}}|\mathbf{S}]$ is called the
imputation distribution. The imputation distribution is typically
specified as a statistical model for the missing data
$\tilde{\mathbf{S}}$ conditioned on the observed data $\mathbf{S}$.

Under the
multiple imputation framework, the distribution
$[\bolds\beta|\mathbf{S}]$ is assumed to
be asymptotically Gaussian. This assumption holds under the
conditions that the joint posterior is unimodal [see, e.g., Chapter~4
of \citet{Gelman2004} for details].
This distribution can then be approximated
using only the posterior predictive mean and variance, which can be
obtained using conditional mean and variance formulae
%
\begin{equation}\label{equ9}
E(\bolds\beta|\mathbf{S}) \approx E_{\tilde{\mathbf{S}}|\mathbf
{S}} \bigl(E(\bolds
\beta|\tilde{\mathbf{S}}) \bigr)
\end{equation}
%
and
%
\begin{equation}\label{equ10}
\operatorname{Var}(\bolds\beta|\mathbf{S}) \approx E_{\tilde{\mathbf{S}}|\mathbf{S}} \bigl(
\operatorname{Var}(\bolds\beta|\tilde{\mathbf{S}}) \bigr)+
\operatorname{Var}_{\tilde{\mathbf{S}}|\mathbf{S}} \bigl( E(\bolds\beta|\tilde{\mathbf{S}})
\bigr).
\end{equation}

If we condition on $\tilde{\mathbf{S}}$, then the posterior distribution
$[\bolds\beta|\tilde{\mathbf{S}}]$ converges
asymptotically to the sampling distribution of the maximum likelihood
estimate (MLE) of $\bolds\beta$ under the likelihood
$[\tilde{\mathbf{S}}|\bolds\beta]$, and we can
approximate $[\bolds\beta|\tilde{\mathbf{S}}]$ by
obtaining the asymptotic sampling distribution of the MLE. This
allows us to use standard maximum likelihood approaches for inference,
which are well developed and computationally efficient for the GLM
formulation in~(\ref{equ6}).

The multiple imputation estimate $\hat{\bolds\beta}_{\mathrm{MI}}$ and
its sampling variance are typically
obtained by approximating the integrals in (\ref{equ9}) and (\ref{equ10}) using a finite
sample from the imputation distribution. The procedure can be
summarized as follows:
\begin{enumerate}[2.]
\item[1.] Draw $K$ different realizations (imputations)
$\tilde{\mathbf{S}}^{(k)} \sim[\tilde{\mathbf{S}}|\mathbf{S}]$
from the path distribution (imputation
distribution). 

\item[2.] For\vspace*{1pt} each realization, find the MLE $\hat{\bolds\beta}^{(k)}$ and
asymptotic variance $\operatorname{Var}(\hat{\bolds\beta}^{(k)})$
of the
estimate under the likelihood
$[\tilde{\mathbf{S}}^{(k)}|\bolds\beta]$ in (\ref{equ6}).

\item[3.] Combine results from different imputations using finite sample
approximations of the conditional
expectation (\ref{equ9}) and variance (\ref{equ10}) results.
\end{enumerate}



This results in point estimates for $E(\bolds\beta|\mathbf{S})$
and $\operatorname{Var}(\bolds\beta|\mathbf{S})$, which can be
used to
construct approximate posterior credible intervals. Combining the multiple
imputation approximation with our GLM formulation of the CTDS
movement model provides a computationally efficient
framework for the statistical analysis of potential drivers of movement.

\subsection{Imputation of continuous-time paths from telemetry data}\label{sec3.2}
Inference using multiple imputation requires the specification of the
imputation distribution $[\tilde{\mathbf{S}}|\mathbf{S}]$, which for
telemetry data is the distribution of the continuous-time movement path
$\tilde{\mathbf{S}}$ conditioned on the observed telemetry data
$\mathbf{S}$. We will consider imputing continuous-time movement
paths by fitting a\break continuous-time movement model to the observations.
Two common\break continuous-time models for movement data are the continuous-time
correlated random walk (CTCRW) of \citet{Johnson2008} and the Brownian
bridge movement model (BBMM) of \citet{Horne2007}. Both assume
continuous movement paths in time and space, and after estimating
model parameters it is straightforward to draw from the posterior
predictive distribution of the continuous-time path
$[\tilde{\mathbf{S}}|\mathbf{S}]$.

The CTCRW model of \citet{Johnson2008} relies on an Ornstein--Uhlenbeck
velocity process.
If the animal's
location and
velocity at an arbitrary time
$t$ are $\mathbf{s}(t)$ and $\mathbf{v}(t)$, respectively, then the
CTCRW model can be specified as follows, ignoring the multivariate
notation for simplicity,
\begin{eqnarray*}
\mathit{d}v(t)&=&\gamma \bigl(\mu-v(t) \bigr)\,\mathit{d}t + \sigma \,dW(t),
\\
s(t)&=&s(0)+\int_0^t v(u)\,du,
\end{eqnarray*}
%
where $\mu$ is a drift term corresponding to long-time scale
directional bias in movement, $\gamma$ controls the rate at which the
animal's velocity reverts to $\mu$, and $\sigma$ scales $W(t)$, which
is standard Brownian motion. This model can
be discretized and formulated as a state-space model, which allows for
efficient estimation of model parameters from telemetry data and
simulation of quasi-continuous discretized paths~$\tilde{\mathbf{S}}$ at
arbitrarily fine time
intervals via the Kalman filter [\citet{Johnson2008a}]. If a
Bayesian framework is
used for inference on $\{\mu,\gamma,\sigma\}$, then \citet
{Johnson2008} show how
to obtain the posterior distribution $[\mu,\gamma,\sigma|\mathbf{S}]$
and\vspace*{2pt} approximate the
posterior predictive distribution of the animal's continuous path
$\tilde{\mathbf{S}}$ using importance
sampling. 

The CTCRW model is a flexible and efficient model for animal
movement that has been
successfully applied to studies of aquatic [\citet{Johnson2008}] and
terrestrial [\citet{Hooten2010b}] animals, and can represent a wide
range of movement behavior, as well as account for location
uncertainty when telemetry locations are observed with error. As
such, we will use the CTCRW model as our primary imputation distribution.
In the supplemental article [\citet{ctdsSupp}], we consider the Brownian
bridge model as an
alternative path imputation distribution and compare it to the CTCRW model.


\subsection{Links to existing methods}\label{sec3.3}

We note that the transition probabilities in (\ref{equ1}) are similar in form
to step selection functions [e.g., \citet{Boyce2002}] in multinomial
logit discrete-choice models for movement data. The key distinction
between the step selection function approach and the approach of
\citet{Hooten2010b} (and, by extension, the approach we present) is the
imputation of the continuous path between telemetry locations.
Imputing the continuous path distribution
allows us to examine movement and resource selection between telemetry
locations, providing a more complete picture of an animal's response
to landscape features and other potential drivers of movement.

The transformation of the movement path from
continuous space to discrete space results in a compression of the data
to a temporal scale that is relevant to the resolution of the
environmental covariates that may be driving movement and selection.
Under the discrete-space, discrete-time dynamic
occupancy approach of
\citet{Hooten2010b}, each discrete-time location is modeled as arising
from a multinomial distribution reflecting transition probabilities
from the animal's location at the previous time. If the animal is in
cell $G_{i_{t-1}}$ at time $t-1$, then define the probability of
transitioning to the $j$th cell at the $t$th time step as $P_{ij_t}$ and
the probability of remaining in cell $i$ as $P_{ii_t}$.
\citet{Hooten2010b} recommend choosing a temporal discretization $\Delta
t$ of the
continuous movement path fine enough to ensure that the animal remains
in each cell for a number of time steps
before transitioning to a neighboring cell.
If an animal is moving slowly relative to the time it takes
to traverse
a grid cell in $\mathbf{G}$, then there will be a long sequence of
locations within
one grid cell before a transition to a neighboring grid cell is made.
In this situation the CTDS approach can be much more efficient than
the discrete-time discrete-space approach of \citet{Hooten2010b}. For
sufficiently small $\Delta t$,
discrete-time transition probabilities are approximated by
$P_{ij_t}\approx\lambda_{i_tj_t} \Delta t$ and $P_{ii_t}\approx
1-\lambda_{i_t} \Delta t$.
Under this model, the probability of the animal remaining in cell
$G_i$ for time equal to $\tau_t$ and then leaving cell $G_i$ is
\[
\lambda_{i_t}\Delta t\prod_{t=1}^{\tau_t/(\Delta t)}P_{ii_t}=
\lambda _{i_t}\Delta tP_{ii}^{\tau_t/\Delta t} =
\lambda_{i_t}\Delta t (1-\lambda_{i_t} \Delta t
)^{\tau
_t/\Delta t}.
\]
Letting $\Delta t \rightarrow0$ results in
%
\begin{equation}\label{equ11}
\lim_{\Delta t \rightarrow0}\lambda_{i_t}\Delta t (1-
\lambda_{i_t} \cdot\Delta t )^{\tau_t/\Delta t} = \lambda_{i_t}\exp
\{-\tau_t \lambda _{i_t} \}.
\end{equation}
Likewise, taking the limit as $\Delta t \rightarrow0$, the
probability of transitioning from cell $G_i$ to $G_k$, given that the
animal is transitioning to some neighboring cell, is
%
\begin{equation}\label{equ12}
\lim_{\Delta t \rightarrow0} \frac{P_{ik_t}}{\sum_j
P_{ij_t}}=\lim_{\Delta t \rightarrow0}
\frac{\lambda_{i_tk_t} \cdot\Delta t}{\lambda_{i_t} \cdot\Delta
t}=\frac{\lambda_{i_tk_t}}{\lambda_{i_t}},
\end{equation}
and (\ref{equ5}) is obtained by multiplying the right-hand sides of (\ref{equ11}) and
(\ref{equ12}). Thus, the CTDS specification could be obtained by using the
sufficient statistics $(\tau_t,\{\lambda_{i_tj_t}\})$ of the
discrete-time, discrete-space approach of \citet{Hooten2010b} in the
limiting case as $\Delta t \rightarrow0$. This data compression is
especially relevant for telemetry data, in which observation windows
can span years or even decades for some animals.

\section{Time-varying behavior and shrinkage estimation}\label{sec4}

In this section we describe how covariate
effects can be allowed to
vary over time using a varying-coefficient model and how variable
selection can be accomplished through regularization.

\subsection{Changing behavior over time}\label{sec4.1}

Animal behavior and response to\break drivers of movement can change
significantly over time. These changes can be driven by external
factors such as changing seasons [e.g., \citet{Grovenburg2009}] or
predator/prey interactions [e.g., \citet{Lima2002}], or by internal
factors such as internal energy levels [e.g., \citet{Nathan2008}].
The most common approach to modeling time-varying behavior in animal
movement is through state switching, typically within a Bayesian
framework
[e.g., \citet
{Morales2004,Jonsen2005,Getz2008,Nathan2008,Forester2009,Gurarie2009,Merrill2010}].
Often, the animal is assumed to exhibit a number of behavioral states,
each characterized by a distinct pattern of movement or response to
drivers of movement. The number of states can be either known and
specified in advance [e.g., \citet{Morales2004,Jonsen2005}] or
allowed to be random [e.g., \citet{Hanks2011PLoS,McClintock2012}].

State-switching models are an intuitive approach to modeling changing
behavior over time, but there are limits to the complexity that can be
modeled using this approach. Allowing the number of states to be
unknown and random requires a Bayesian approach with a changing
parameter space. This is typically implemented using reversible-jump
MCMC methods [e.g., \citet{Green1995,McClintock2012,Hanks2011PLoS}],
which are computationally expensive and can be difficult to tune. Our
approach is to use a computationally efficient GLM (\ref{equ6}) to analyze
parameters related to drivers of animal movement. Instead of using
the common state-space approach, we employ varying-coefficient models
[e.g., \citet{Hastie1993}] to model time-varying behavior in animal
movement. A similar approach to modeling time-varying behavior in
animal movement was taken by \citet{Breed2012}.

For simplicity in notation, consider the case where there is only one
covariate $x$ in the model (\ref{equ1}) and no intercept term. The model for
the transition rate will typically contain an intercept
term and multiple covariates $\{x\}$, and the varying-coefficient
approach we present generalizes easily to this case. In a
time-varying coefficient model, we allow the parameter $\beta(t)$ to
vary over time in a functional (continuous) fashion. The transition
rate (\ref{equ1}) then becomes
\[
\lambda_{i_tj_t}\bigl(\beta(t)\bigr)=\exp \bigl\{x_{i_tj_t}\beta(t)
\bigr\},
\]
where $t$ is the time of the observation and $x_{ij}$ is the value of
the covariate related to the exponential rate of moving from cell $i$ to
cell $j$. We model the functional regressor
$\beta(t)$ as a linear combination of $n_{\mathrm{spl}}$ spline basis
functions $\{\bolds\phi_k(t),k=1,\ldots,n_{\mathrm{spl}}\}$:
\[
\beta(t)=\sum_{k=1}^{n_{\mathrm{spl}}}
\alpha_k \bolds\phi_k(t).
\]
%

Under this varying-coefficient specification, (\ref{equ1}) can be rewritten as
%
\begin{eqnarray}\label{equ13}
\lambda_{i_tj_t}&=&\exp \bigl\{x_{i_tj_t}\beta(t) \bigr\}
\nonumber
\\
&=&\exp \Biggl\{x_{i_tj_t}\sum_{k=1}^{n_{\mathrm{spl}}}
\alpha_k \bolds\phi_k(t) \Biggr\}
\\
&=&\exp \bigl\{\bolds\psi_{i_tj_t}' \bolds\alpha
\bigr\},\nonumber
\end{eqnarray}
where\vspace*{1pt} $\bolds\alpha=(\alpha_1,\ldots,\alpha_{n_{\mathrm{spl}}})'$ and
$\bolds\psi_{i_tj_t}=x_{i_tj_t}\cdot(\phi_1(t),\ldots,\phi_{n_{\mathrm{spl}}}(t))'$.
The result is that the varying-coefficient model can be represented by a
GLM with a modified design matrix. This
specification provides a flexible framework for allowing the effect of
a driver of movement ($x$) to vary over time that is computationally
efficient and simple to implement
using standard GLM software. For our asymptotic arguments in Section~\ref{sec3.1} to hold, we will only consider the case where $n_{\mathrm{spl}}$ is fixed
and the temporal variation in the $\beta(t)$ models periodic (e.g.,
diurnal) changes in movement behavior.

\subsection{Regularization}\label{sec4.2}

The model we have specified is likely to be overparameterized,
especially if we utilize a varying-coefficient model (\ref{equ13}).
Animal movement behavior is complex, and a typical study could
entail a large number of \mbox{potential} drivers of movement, but an
animal's response to each of those drivers of
movement is likely to change over time, with only a few drivers being
relevant at any one time. Under these assumptions, many of the
parameters $\alpha_k$ in (\ref{equ13}) are likely to be very small or zero.
Multicollinearity is also a potential problem, as many potential
drivers of movement could be correlated with each other.

The most common approach to these issues is penalization or
regularization [e.g., \citet{Tibshirani1996,HootenHobbs2014}]. We
propose a shrinkage estimator of $\bolds\alpha$ using a
lasso penalty [\citet{Tibshirani1996}]. The typical maximum likelihood
estimate of
$\bolds\alpha$ is obtained by maximizing the likelihood
$[\mathbf{Z},\bolds\tau|\bolds\alpha]$ from (\ref{equ6}) or,
equivalently, by maximizing the log-likelihood $\log
[\mathbf{Z},\bolds\tau|\bolds\alpha]$. The lasso estimate
is obtained by maximizing the penalized log-likelihood, where the
penalty is proportional to the sum of the absolute values of the
regression parameters $\{\alpha_k\}$:
%
\begin{equation}\label{equ14}
\hat{\bolds\alpha}_{\mathrm{lasso}}=\max_{\bolds\alpha} \Biggl\{
\log [\mathbf{Z},\bolds\tau|\bolds\alpha] -\gamma\sum
_{k=1}^K |\alpha_k| \Biggr\}.
\end{equation}
As the tuning parameter $\gamma$ increases, the absolute values of the
regression parameters $\{\alpha_k\}$ are ``shrunk'' to zero, with the
parameters that best describe the variation in the data being shrunk
more slowly than parameters that do not. Cross-validation is
typically used to set the tuning parameter $\gamma$ at a level that
optimizes the model's predictive power.

Shrinkage approaches such as the lasso are well developed for GLMs,
and computationally-efficient methods are available for fitting GLMs
to data [e.g., \citet{Friedman2010}]. Recent work has also applied the
lasso to
multiple imputation estimators [e.g., \citet{Chen2011}]. The main challenge
in applying the lasso to multiple imputation is that a parameter may
be shrunk to zero in the analysis of one imputation but not in the
analysis of another. If the lasso is used for variable
selection, a group lasso penalty
[\citet{Yuan2006}] can be specified
in which a group of parameters is constrained to either all equal zero
or all be nonzero together. In the case of multiple imputation, we
consider the joint analysis of all imputations and constrain the set
of $\{\alpha_p^{(k)},k=1,\ldots,K\}$, where $p$ indexes the parameters
in the model and $k$ indexes the imputations, to either all equal zero
or all be nonzero together. This group lasso sets the requirement
that a parameter must either be zero for all imputations or nonzero for all
imputations. One simple approach to implementing this group lasso is
to combine all imputations and analyze the aggregate paths as if they
were independent observed paths. This amounts to the stacked lasso
estimate of \citet{Chen2011} and is reminiscent of data cloning
[\citet{Lele2011}]. We note that this approach does not yield
straightforward estimates of the uncertainty about the lasso
estimates. We will focus on a full Bayesian analysis with lasso prior
to characterize the uncertainty in $\bolds\alpha$ under a lasso approach.

In a full Bayesian analysis we consider specifying a shrinkage prior
distribution on
$\bolds\alpha$ such that the posterior mode of
$\bolds\alpha|\mathbf{S}$ is identical to the lasso estimate
(\ref{equ14}). Instead of the Gaussian prior in (\ref{equ7}), we follow
\citet{Park2008} and consider a hierarchical prior
specification:
%
\begin{equation}\label{equ15}
\alpha_k|\sigma^2_k \sim N\bigl(0,
\sigma^2_k\bigr),\qquad k=1,\ldots,K,
\end{equation}
where the prior on $\sigma^2_k$ is conditioned on the shrinkage
parameter $\gamma$:
%
\begin{equation}\label{equ16}
\bigl[\sigma^2_k|\gamma^2\bigr] \propto
\gamma^2 \exp\bigl\{-\gamma^2 \sigma^2_k/2
\bigr\},\qquad k=1,\ldots,K.
\end{equation}
Then, marginalizing over the $\sigma_k^2$ gives a Laplace prior distribution
on $\alpha$ conditioned only on $\gamma$:
\begin{eqnarray*}
[\alpha_k|\gamma]&=&\int_0^\infty
\bigl[\alpha_k|\sigma^2_k\bigr] \bigl[
\sigma^2_k|\gamma\bigr]\,\mathit{d}\sigma^2_k
\\
&\propto&\int_0^\infty\frac{1}{\sqrt{2\pi\sigma_k^2}} \exp\bigl
\{-\alpha^2_k/\bigl(2\sigma_k^2
\bigr)\bigr\}\gamma^2 \exp\bigl\{-\gamma^2
\sigma^2_k/2\bigr\}\,\mathit{d}\sigma^2_k
\\
&=&\frac{\gamma}{2}\exp\bigl\{-\gamma|\alpha_k|\bigr\},
\end{eqnarray*}
where the last step uses the representation of the Laplace
distribution as a scale mixture of Gaussian random variables with
exponential mixing density [e.g., \citet{Park2008}].
Maximizing the resulting log-posterior predictive distribution for
$\bolds\alpha$ gives us the lasso estimate (\ref{equ14}).

The hyperparameter $\gamma$ controls the amount of shrinkage in the
Bayesian lasso. While a prior distribution could be assigned to
$\gamma$, we take an empirical approach and estimate $\gamma$
using cross-validation in the penalized likelihood approach (\ref{equ14}) to the
lasso. This estimate can then be used to set the value of the
hyperparameter $\gamma$ in the Bayesian lasso analysis.

\section{Drivers of animal movement}\label{sec5}

We now provide some examples showing how a range of hypothesized
drivers of movement could be modeled within the CTDS framework. We
consider two distinct categories for drivers of movement from cell
$G_i$ to cell $G_j$: location-based drivers
($\{p_{ki},k=1,2,\ldots,\break K\}$), which are determined only by the
characteristics of cell $G_i$, and directional drivers
($\{q_{lij},l=1,2,\ldots,L\}$), which vary with direction of movement.
Under a time-varying coefficient model for each driver, the transition
rate (\ref{equ1}) from cell $G_i$ to cell $G_j$ is
%
\begin{equation}\label{equ17}
\lambda_{ij} \bigl(\bolds\beta(t) \bigr)=\exp \Biggl\{
\beta_0(t)+\sum_{k=1}^{K}p_{ki}
\beta_k(t)+\sum_{l=1}^{L}q_{lij}
\beta_l(t) \Biggr\},
\end{equation}
where $\beta_0(t)$ is a time-varying intercept term, $\{\beta_k(t)\}$
are time-varying effects related to location-based drivers of movement, and
$\{\beta_l(t)\}$ are time-varying effects related to directional drivers
of movement.
We consider both location-based and directional drivers in what follows.

\subsection{Location-based drivers of movement}\label{sec5.1}

Location-based drivers of
movement can be used to examine differences in animal movement rates
that can be explained by the environment an animal resides in. For
example, if the animal is in a patch of highly desirable terrain,
surrounded by less-desirable terrain, a
location-based driver of movement could be used to model the animal's
propensity to stay in the desirable patch and move quickly through
undesirable terrain.
In the CTDS
context, location-based
drivers would be covariates dependent only on the characteristics of
the cell where the animal is currently located. Large positive
(negative) values of the corresponding $\beta_k(t)$
would indicate that the animal tends to transition quickly (slowly)
from a cell containing the cover type in question.

\subsection{Directional bias in movement}\label{sec5.2}

In contrast to location-based drivers, which describe the effect that the
local environment has on movement rates,
directional drivers of movement [\citet
{Brillinger2001,Hooten2010b,Hanks2011PLoS}] capture
directional bias in movement patterns.

\begin{figure}

\includegraphics{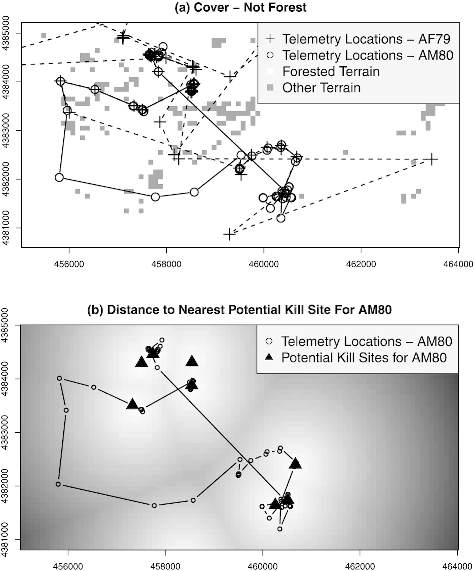}

\caption{Telemetry data for a female mountain lion (AF79) and her male
cub (AM80). A location-based covariate was defined by landcover that was
not predominanty forested \textup{(a)}. Potential kill sites were
identified, and a directional (bias)
covariate defined by a vector pointing toward the closest kill site \textup{(b)}
was also used in the CTDS model.}\label{fig2}
\end{figure}

A directional driver of movement (or bias effect in our GLM) is defined
by a vector which points
toward (or away) from something that is hypothesized to attract (or
repel) the animal in question.
Let $\mathbf{v}_l$ be the vector corresponding to the $l$th directional
driver of movement. In the CTDS model for animal movement, the animal
can only transition from cell $G_i$ to one of its neighbors
$G_j\dvtx j\sim i$. Let $\mathbf{w}_{ij}$ be a unit vector pointing from
the center of cell $G_i$ in the direction of the center of cell
$G_j$. Then the covariate $q_{lij}$ relating the $l$th directional
driver of movement to\vadjust{\goodbreak} the transition rate from cell $G_i$ to cell
$G_j$ is the inner product of $\mathbf{v}_l$ and $\mathbf{w}_{ij}$:
\[
q_{lij}=\mathbf{v}_l'\mathbf{w}_{ij}.
\]
Then $p_{lij}$ will be positive when $\mathbf{v}_l$ points
nearly in the direction of cell $G_j$, negative when $\mathbf{v}_l$ points
directly away from cell $G_j$, and zero if $\mathbf{v}_l$ is
perpendicular to the direction from cell $G_i$ to cell $G_j$.

\section{Example: Mountain lions in Colorado}\label{sec6}

We illustrate our CTDS random walk approach to modeling animal
movement through a study of mountain lions (\emph{Puma concolor}) in
Colorado, USA. R code to download all needed files and
replicate this analysis is available from the R-forge website
(\surl{http://r-forge.r-project.org/projects/ctds/}).
As part
of a larger study, a female mountain lion, designated AF79, and her
subadult cub, designated AM80, were fitted with global positioning
system (GPS)
collars set to transmit location data every 3 hours. We analyze the
location data $\mathbf{S}$
from two weeks (14 days) of location information for these two
animals (Figure~\ref{fig2}).

We fit the CTCRW model of \citet{Johnson2008} to both animals' location
data using the ``crawl'' package
[\citet{crawl}] in the R statistical computing environment [\citet{R}].


For covariate data, we used a landcover map of the state of Colorado
created by the Colorado Vegetation Classification Project
(\url{http://ndis.nrel.colostate.edu/coveg/}), which is a joint project of
the Bureau of
Land Management and the Colorado Division of Wildlife. The landcover map
contained gridded landcover information at 100~m square resolution.
The area traveled by the
two animals in our study was predominantly forested. To assess how
the animals' movement differed when in terrain other than forest, we
created an indicator covariate where all forested grid cells were
assigned a value of zero, and all cells containing other cover types,
including developed land, bare
ground, grassland and shrubby terrain, were assigned a value of one
[Figure~\ref{fig2}(a)]. This covariate was used as a
location-based covariate in the CTDS model.


For the subadult male AM80, we created a set of potential kill sites
(PKS) by
examining the original GPS location data [Figure~\ref{fig2}(b)]. \citet{Knopff2009}
classified a
location as a PKS if two or more GPS locations were found within 200~m
of the site within a six-day period. We added an additional constraint
that at least one of the GPS locations be during nighttime hours (9~pm
to 6~am) for the point to be classified a PKS.
We then created a covariate raster
layer containing the distance to the nearest PKS
for each grid
cell [Figure~\ref{fig2}(b)]. A directional covariate defined by a vector pointing
toward the nearest PKS was included in the CTDS model.

To examine how the movement path of the mother AF79 affected the movement
path of the cub AM80, we included a directional covariate in the
CTDS model for AM80 defined
by a vector pointing from the cub's location to the
mother's location at each time point.

We also included a directional covariate pointing in
the direction of the most recent movement at each time point. This
covariate measures the strength of correlation between moves and thus
the strength of the directional persistence shown by the animal's
discrete-space movement path.
The CTCRW imputation distribution assumes an underlying correlated
movement model,
while the Brownian bridge model does not. See the online supplement
for details [\citet{ctdsSupp}].

\subsection{Comparison of methods under time-homogeneous model}\label{sec6.1}
We first compare a full Bayesian analysis of the path of AM80 to the
multiple imputation approximation to the
posterior mean (\ref{equ9}) and variance (\ref{equ10}). For this first analysis, we do not
assume any time-varying behavior, but rather model the cub's mean
response over time to the landscape, identified PKSs and the movement
path of AF79. For both the full Bayesian analysis and the multiple
imputation approximations we used the CTCRW imputation distribution.
We used a Markov chain Monte Carlo algorithm to draw
20,000 samples from the posterior predictive distribution of
$\bolds\beta|\mathbf{S}$ for AM80. We discarded the first 5000
as burn-in and used the remaining samples to approximate the posterior
predictive distribution. Posterior means and standard deviations are
shown in Table~\ref{tab1}. Each parameter whose posterior predictive
distribution's 95$\%$ equal-tailed credible interval does not overlap
zero is marked with a star in Table~\ref{tab1}. We then applied the multiple imputation
approach to approximate the posterior
distribution using\vspace*{1pt} the $K=2, 5, 10$ and 50 continuous paths drawn from
the CTCRW
imputation distribution: $[\tilde{\mathbf{S}}|\mathbf{S}]$. The
resulting mean and posterior standard deviations are given in Table~\ref{tab1}. We constructed symmetric asymptotically normal 95$\%$ confidence
intervals for each regression parameter, and mark each estimate with a
star in Table~\ref{tab1} when the confidence interval does not overlap zero.
The multiple imputation results
approximate the mean and variance of the posterior
predictive distribution in this example with reasonable precision,
even when very few imputations are used, and when $K=50$ imputed paths
are used, the multiple imputation approximation yields results that
are nearly identical to the results from the fully Bayesian analysis.

\begin{table}
\tabcolsep=5pt
\caption{Results on regression parameters related to movement
behavior. Entries are Bayesian posterior predictive means ($\hat{\beta
}$) and
standard deviations (s.e.) for the fully Bayesian analysis (Bayes),
and multiple imputation approximations to the same for the multiple
imputation analyses. Results are shown for varying numbers of
imputations $K$ from the continuous-time correlated random walk
(CTCRW) path
imputation distribution $[\tilde{\mathbf{S}}|\mathbf{S}]$.
Starred
entries indicate parameters with a 95$\%$ Bayesian credible interval
that does not overlap zero}\label{tab1}
\begin{tabular*}{\tablewidth}{@{\extracolsep{\fill}}@{}lcccccccccc@{}}
\hline
&&& \multicolumn{2}{c}{\textbf{Forest cover}} & \multicolumn{2}{c}{\textbf{Dist. to PKS}} & \multicolumn{2}{c}{\textbf{Dist. to AF79}} & \multicolumn{2}{c@{}}{\textbf{CRW}}\\[-6pt]
&&& \multicolumn{2}{c}{\hrulefill} & \multicolumn{2}{c}{\hrulefill} & \multicolumn{2}{c}{\hrulefill} & \multicolumn{2}{c@{}}{\hrulefill}\\
\textbf{Method} & $\bolds{[\tilde{\mathbf{S}}|\mathbf{S}]}$ & $\bolds{K}$
& \multicolumn{1}{c}{$\bolds{\hat{\beta}}$} & \multicolumn{1}{c}{\textbf{s.e.}}
& \multicolumn{1}{c}{$\bolds{\hat{\beta}}$} & \multicolumn{1}{c}{\textbf{s.e.}}
& \multicolumn{1}{c}{$\bolds{\hat{\beta}}$} & \multicolumn{1}{c}{\textbf{s.e.}}
& \multicolumn{1}{c}{$\bolds{\hat{\beta}}$} & \multicolumn{1}{c}{\textbf{s.e.}}\\
\hline
Bayes& CTCRW& NA & 0.326 & 0.197 & 0.297$^*$ & 0.043 & 0.059 & 0.048 & 0.398$^*$ & 0.0518 \\
MI & CTCRW&50 & 0.326 & 0.197 & 0.297$^*$ & 0.043 & 0.059 & 0.048 & 0.398$^*$ & 0.051\phantom{0} \\
MI & CTCRW&10 & 0.334 & 0.197 & 0.305$^*$ & 0.042 & 0.063 & 0.050 & 0.399$^*$ & 0.0487 \\
MI & CTCRW&\phantom{0}5 & 0.372 & 0.154 & 0.293$^*$ & 0.040 & 0.076 & 0.061& 0.407$^*$ & 0.043\phantom{0} \\
MI & CTCRW&\phantom{0}2 & 0.228 & 0.168 & 0.300$^*$ & 0.046 & 0.035 & 0.055 & 0.431$^*$ & 0.040\phantom{0} \\
\hline
\end{tabular*}
\end{table}

The results show that much of the subadult male's movement can be
explained by a correlated random walk with attractive
points at PKSs [Figure~\ref{fig2}(b)]. The results also show that the animal's movement
behavior is fairly homogeneous when in forested and in nonforested
terrain. These results are consistent for all approaches using the
CTCRW imputation distribution.

\subsection{Simulation study}\label{sec6.2}


\begin{table}
\tabcolsep=5pt
\caption{Simulation study results. A simulation study was conducted,
by setting the true covariate effects for
``Not forest,'' ``Direction to nearest PKS'' and ``Distance to
AF79'' to various values motivated by the estimates in Section~\protect\ref{sec6.1}.
We then simulated a CTDS random walk under the true parameters, and
thinned the simulated path to ``observed'' locations at
four-hour intervals (to simulate regular telemetry observations).
The resulting simulated observations were fit
using our proposed approach using the CTCRW model to impute continuous-time
paths and a lasso penalty on the fitted GLM. This simulation study
was repeated for the case when the true covariate effects are all
zero. In each case, 1000 paths were simulated and fit, with the
results summarized below}\label{tab2}
\begin{tabular*}{\tablewidth}{@{\extracolsep{\fill}}@{}lccccc@{}}
\hline
& \textbf{True} & \textbf{Proportion}         & \textbf{Proportion}     &  &  \\
\textbf{Covariate}                    & \textbf{value} & $\bolds{\hat{\beta}\neq0}$ & $\bolds{\hat{\beta}=0}$ & \textbf{Min} & \textbf{Max} \\
\hline
Not forest & 0.000 & 0.000 & 1.000 & 0.000 & 0.000 \\
Direction to PKS & 0.300 & 0.634 & 0.866 & 0.000 & 0.217 \\
Distance to AF79 & 0.000 & 0.000 & 1.000 & 0.000 & 0.000
\\[3pt]
Not forest & 0.000 & 0.000 & 1.000 & 0.000 & 0.000 \\
Direction to PKS & 0.000 & 0.002 & 0.998 & 0.000 & 0.016 \\
Distance to AF79 & 0.000 & 0.000 & 1.000 & 0.000 & 0.000 \\
\hline
\end{tabular*}
\end{table}

We conducted a simulation study motivated by our data analysis to
examine our ability to find the correct subset model using multiple
imputation with lasso penalty.
We are interested in identifying which
parameters affect animal movement and directional bias, and so focus
on a group lasso penalty which will force estimates for regression
parameters to be either zero or nonzero in all imputations. An
alternative approach would be to obtain a lasso estimate of the
regression parameters (\ref{equ14}) for each imputated path, and then combine
them using the standard combining rules.

We first simulated a CTDS movement path using the forest cover and
direction to nearest PKS covariates from our mountain lion analysis,
as well as a simulated covariate meant to mimic the directional effect
of the conspecific (AF79).
Various combinations of true parameter values were specified, and a
full CTDS path was simulated for a two-week period (equal to the
observation period of the mountain lions in our study). We then
simulated telemetry data from the CTDS path by recording the simulated
location only every four hours. The resulting simulated telemetry
locations were used to estimate the movement parameters using our
approach with a CTCRW imputation distribution and lasso penalty, with
the lasso tuning parameter chosen by 10-fold cross-validation using
the ``glmnet'' package [\citet{Friedman2010}] in R. This was repeated
1000 times for each set of parameters. The results are shown in Table~\ref{tab2}.

Our approach is very accurate at estimating model parameters as equal
to zero when the
true parameter is zero. When the true value of the parameter related
to the directional gradient toward the nearest PKS is positive (0.30),
the approach correctly estimates this parameter as positive 86.6$\%$ of
the time, and never incorrectly estimates this parameter as being
negative.

From this simulation study we see that our proposed approach with
lasso penalty
provides a conservative estimate of the relationship between an
animal's observed movement and the potential drivers of animal
movement in the model (\ref{equ17}).

\subsection{Time-varying behavior}\label{sec6.3}

We next examine changing movement behavior over time using a varying-coefficient
model for each covariate in the model, where behavior was allowed to
vary with time of day. For all covariates we
specified a \mbox{B-}spline basis expansion with regularly-spaced spline
knots at 6 hour intervals over the course of a 24 hour period.
Observations over multiple days (14 days in this study) are
replications in this model and allow for inference about diurnal
changes in movement
behavior.

\begin{figure}

\includegraphics{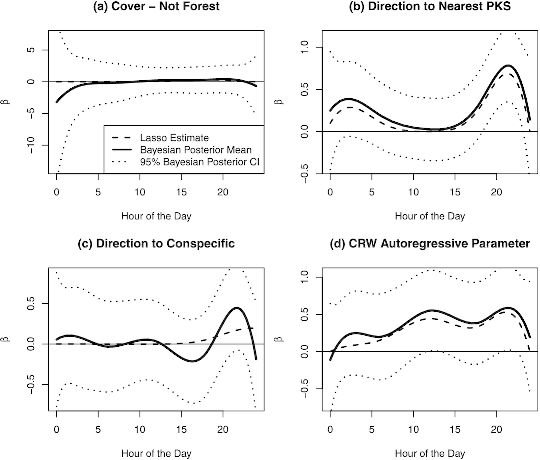}

\caption{Time-varying results for the location-based and directional
covariates in the conti\-nuous-time discrete-space
model for a male mountain lion (AM80) obtained using a lasso
shrinkage prior. The $x$-axis is time of day in hours. The $y$-axis is
the effect size.}\label{fig3}\vspace*{-3pt}
\end{figure}

For this analysis, we fit the CTDS model with CTCRW imputation
distribution and a lasso penalty. After estimating the model
parameters and choosing the lasso tuning parameter using
cross-validation, we used the chosen lasso tuning parameter $\gamma$
as a hyperparameter in the full Bayesian model with lasso shrinkage
prior (\ref{equ15})--(\ref{equ16}). The resulting posterior predictive mean and
equal-tailed $95\%$ credible interval bounds for $\bolds\beta(t)$
are shown in Figure~\ref{fig3}.


In Figure~\ref{fig3}(b) the peak in the value of the $\beta(t)$ associated with
movement toward the nearest PKS indicates the animal shows some
preference for returning to a PKS near dusk (8~pm).
The confidence bands of the
other parameters include zero throughout the day, indicating that we
lack evidence that the animal's response to the relevent covariates is
synchronous with the diurnal cycle.

\section{Discussion}\label{sec7}

While we have couched our CTDS approach in terms of modeling animal
movement, we can also view this approach in terms of resource
selection [e.g., \citet{Manly2002}]. \citet{Johnson2008} describe a
general framework for the analysis of resource selection from telemetry
data using a weighted distribution approach where an observed
distribution of resource use is seen as a reweighted version of a
distribution of available resources,\vadjust{\goodbreak} and the resource selection
function (RSF) defines the preferential use of resources by the
animal. \citet{Warton2010} describe a point process approach to
resource selection that can be fit using a Poisson GLM, similar to the
CTDS model we describe here. In the context of \citet{Warton2010}, the
CTDS approach can be viewed as a resource selection analysis with the
available resources at any time defined as the neighboring grid
cells. The transition rate (\ref{equ17}) of the CTDS process to each
neighboring cell contains information about the availability of each
cell, as well as the RSF that defines preferential use of the resources in
each cell.

One alternative to our continuous time model for animal movement is
the spatio-temporal point process modeling approach of
\citet{JohnsonJAE2013}, where the movement process is considered as a
set of points that exist in space
and time, instead of as a dynamic process occurring in space and time.
In the spatio-temporal point process context,
telemetry points can arise in a space that is both geographical and
temporal, and \citet{JohnsonJAE2013} integrate over the temporal dimension
and arrive at a marginal spatial point process model. Our approach is
explicitly dynamic in that it models actual transition probabilities
as function spatio-temporally varying environmental and ecological
conditions. Furthermore, we allow for additional flexibility and
predictive ability in our approach through the use of regularization.

Representing a CTMC model for CTDS animal movement in terms of a
Poisson GLM likelihood (\ref{equ6}) allows for the possibility of inference
under a wide variety of statistical approaches. An alternative to
our Bayesian approach based on MCMC, generalized additive modeling
approaches and software [e.g., \citet{Wood2011}], as well as approximate
Bayesian approaches such as integrated nested
Laplace approximations [INLA, \citet{Rue2009}], could be used for
inference on time-varying parameters in (\ref{equ13}).

The use of directional drivers of movement has a long history.
\citet{Brillinger2001} model animal movement as a continuous-time,
continuous-space random walk where
the drift term is the gradient of a ``potential function''
that defines an animal's external drivers of movement.
\citet{Tracey2005} use circular distributions to model how an animal
moves in response to a vector pointing toward an object that may
attract or repel the animal. \citet{Hanks2011PLoS} and \citet
{McClintock2012} make extensive use of gradients to model
directed movements and movements about a central location.
In our
study of mountain lion movement data, we used directional drivers of
movement to model conspecific interaction between a mother (AF79) and her
cub (AM80). Interactions between predators and prey could also be
modeled using directional covariates defined by vectors pointing between
animals. Some movements based on memory could also be modeled using
directional covariates. For example, a directional covariate defined by a
vector pointing to the animal's location\vadjust{\goodbreak} one year prior could be used to
model seasonal migratory behavior. The ability to model both location-based
and directional drivers of movement make the CTDS framework a flexible and
extensible framework for modeling complex behavior in animal movement.


\section*{Acknowledgments}
We would like to thank Jake Ivan and multiple anonymous reviewers for
providing valuable feedback on this manuscript. Funding for this
project was provided by Colorado Parks and Wildlife
(\#1201). Any use of trade, firm or product names is for descriptive
purposes only and does not imply endorsement by the U.S. Government.\vspace*{-3pt}

%
\begin{supplement}
\stitle{Alternate path imputation distribution}
\slink[doi]{10.1214/14-AOAS803SUPP}
\sdatatype{.pdf}
\sfilename{aoas803\_supp.pdf}
\sdescription{This supplement contains details of a Brownian bridge
path imputation distribution and its use with our CTDS approach to
modeling animal movement.}\vspace*{-3pt}
\end{supplement}



%

\printaddresses
\end{document}